\documentclass{llncs}
\usepackage{epsfig,amsmath,amssymb,url}

\title{
    Inferring AS Relationships: \\
    Dead End or Lively Beginning?
}

\author{ Xenofontas Dimitropoulos\inst{1,2} \and Dmitri Krioukov\inst{2} 
	\and Bradley Huffaker\inst{2} \and \\ kc claffy\inst{2} \and George Riley\inst{1}}
\institute{School of Electrical and Computer Engineering\\
	Georgia Institute of Technology\\
	Atlanta, Georgia 30332--0250\\
	\email{fontas@ece.gatech.edu,riley@ece.gatech.edu}
\and
  Cooperative Association for Internet Data Analysis (CAIDA)\\
  La Jolla, California 92093--0505\\
  \email{dima@caida.org,brad@caida.org,kc@caida.org}
}

\begin{document}

\maketitle

\begin{abstract}

Recent techniques for inferring business relationships
between ASs~\cite{DiBaPaPi03,ErHaSch02} have yielded
maps that have extremely few {\em invalid\/} BGP paths in the
terminology of Gao~\cite{Gao01}.  However, some
relationships inferred by these newer algorithms are
incorrect, leading to the deduction of unrealistic AS hierarchies. We
investigate this problem and discover what causes it. Having
obtained such insight, we generalize the problem of AS
relationship inference as a multiobjective optimization
problem with node-degree-based corrections to the original
objective function of minimizing the number of invalid paths.
We solve the generalized version of the problem using the
semidefinite programming relaxation of the MAX2SAT problem.
Keeping the number of invalid paths small, we obtain a
more veracious solution than that yielded by recent heuristics.

\end{abstract}

\section{Introduction}

As packets flow in the Internet, money also flows, not
necessarily in the same direction.  Business relationships
between ASs reflect both flows, indicating a direction of
money transfer as well as a set of constraints to the flow
of traffic. Knowing AS business relationships is therefore
of critical importance to providers, vendors, researchers,
and policy makers, since such knowledge sheds more light on
the relative ``importance'' of ASs.

The problem is also of multidimensional interest to the research
community. Indeed, the Internet AS-level topology and its
evolutionary dynamics result from business decisions
among Internet players. Knowledge of AS relationships in the
Internet provides a valuable validation framework for
economy-based Internet topology evolution modeling, which in
turn promotes deeper understanding of the fundamental laws
driving the evolution of the Internet topology and its hierarchy.

Unfortunately, the work on inferring AS relationships from BGP
data has recently encountered difficulties. We briefly
describe this situation in its historical context.

Gao introduces the AS relationship inference problem in her
pioneering paper~\cite{Gao01}. This work approximates reality by
assuming that any AS-link is of one of the following three types:
customer-provider, peering, or sibling. If all ASs strictly adhere
to import and export policies described in~\cite{Gao01}, then
every BGP path must comply with the following hierarchical pattern:
an uphill segment of zero or more customer-to-provider or
sibling-to-sibling links, followed by zero or one peer-to-peer
links, followed by a downhill segment of zero or more
provider-to-customer or sibling-to-sibling links. Paths with the
described hierarchical structure are deemed {\em valid}. After
introducing insight about valid paths, Gao proposes an inference
heuristic that identifies top providers and peering links based 
on AS degrees and valid paths.

In~\cite{SuAgReKa02}, Subramanian {\it et al}.~(SARK) slightly
relax the problem by not inferring sibling links, and introduce a
more consistent and elegant mathematical formulation. The authors
render the problem into a combinatorial optimization problem:
given an undirected graph~$G$ derived from a set of BGP paths~$P$,
assign the edge type (customer-provider or peering) to every edge
in~$G$ such that the total number of valid paths in~$P$ is
maximized. The authors call the problem the
type-of-relationship~(ToR) problem, conjecture that it is
NP-complete, and provide a simple heuristic approximation.

Di~Battista {\it et al}.~(DPP) in~\cite{DiBaPaPi03} and
independently Erlebach {\it et al}.~(EHS) in~\cite{ErHaSch02}
prove that the ToR problem is indeed NP-complete. EHS prove also
that it is even harder, specifically APX-complete.\footnote{There exists no
polynomial-time algorithm approximating an APX-complete problem
above a certain {\em inapproximability\/} limit (ratio) dependent
on the particular problem.} More importantly for practical
purposes, both DPP and EHS make the straightforward observation
that peering edges {\em cannot\/} be inferred in the ToR problem
formulation. Indeed, as the validation data presented by 
Xia {\it et al}. in~\cite{XiaGao04} indicates, only 24.63\%
of the validated SARK peering links are correct.

Even more problematic is the following dilemma. DPP (and
EHS) come up with heuristics that outperform the SARK
algorithm in terms of producing smaller numbers of
invalid paths~\cite{DiBaPaPi03,ErHaSch02}. Although these results
seem to be a positive
sign, closer examination of the AS relationships produced by the
DPP algorithm~\cite{rimondini02} reveals that the DPP inferences
are further from reality than the SARK inferences. In the next
section we show that improved solutions to the ToR problem do not
yield practically correct answers and contain obviously
misidentified edges, e.g.~well-known global providers appear as
customers of small ASs. As a consequence, we claim that improved
solutions to the unmodified ToR problem do not produce realistic
results.

An alternative approach to AS relationship inference is to
disregard BGP paths and switch attention to other data sources
(e.g.~WHOIS)~\cite{SiFa04,HuLeOdWa04}, but nothing suggests that
we have exhausted all possibilities of extracting relevant
information from BGP data. Indeed, in this study we seek to answer
the following question: can we adjust the original (ToR) problem
formulation, so that an algorithmic solution to the modified
problem would yield a better answer from the practical
perspective?

The main contribution of this paper is that we positively answer
this question. We describe our approach and preliminary results in
the subsequent two sections, and conclude by describing
future directions of this work.

\section{Methodology}

\subsection{Inspiration behind our approach}

The main idea behind our approach is to formalize our knowledge
regarding why improved solutions to the ToR problem fail to yield
practically right answers. To this end we reformulate the ToR
problem as a multiobjective optimization problem introducing
certain corrections to the original objective function.
We seek a modification of the original objective function,
such that the minimum of the new objective function reflects an AS
relationship mapping that is closer to reality.

\subsection{Mapping to 2SAT}

To achieve this purpose, we start with the DPP and EHS
results~\cite{DiBaPaPi03,ErHaSch02} that deliver the fewest
invalid paths. Suppose we have a set of BGP paths~$P$ from which
we can extract the undirected AS-level graph~$G(V,E)$. We
introduce {\em direction\/} to every edge in~$E$ {\em from\/} the
customer AS {\em to\/} the provider AS. Directing edges in~$E$
induces direction of edges in~$P$. A path in~$P$ is valid if it
does not contain the following invalid pattern: a provider-to-customer edge
followed by a customer-to-provider edge. The ToR problem is to
assign direction to edges in~$E$ minimizing the number of paths
in~$P$ containing the invalid pattern.

The problem of identifying the directions of all
edges in~$E$ making {\em all\/} paths in~$P$ valid---assuming
such edge orientation exists---can be reduced to the 2SAT
problem.\footnote{2SAT is a
variation of the satisfiability problem: given a set of clauses
with two boolean variables per clause~\mbox{$l_i \vee l_j$}, find
an assignment of values to variables satisfying all the clauses.
MAX2SAT is a related problem: find the assignment maximizing the
number of simultaneously satisfied clauses.}  Initially,
we arbitrarily direct all edges in~$E$ and introduce a boolean
variable~$x_i$ for every edge~$i$, $i=1\ldots\big|E\big|$. If the
algorithms described below assign the value {\em true\/} to~$x_i$,
then edge~$i$ keeps its original direction, while assignment of
{\em false\/} to~$x_i$ reverses the direction of~$i$. We then
split each path in~$P$ into pairs of adjacent edges involving
triplets of ASs (all 1-link paths are always valid) and perform
mapping between the obtained pairs and 2-variable clauses as shown
in~Table~\ref{table:mapping}. The mapping is such that only
clauses corresponding to the invalid path pattern yield the {\em
false\/} value when both variables are {\em true}. If there exists
an assignment of values to all the variables such that all clauses
are satisfied, then this assignment makes all paths valid.

To solve the 2SAT problem, we construct a dual graph, the 2SAT
graph $G_{2SAT}(V_{2SAT},E_{2SAT})$, according to the rules shown
in~Table~\ref{table:mapping}: every edge~\mbox{$i \in E$} in the
original graph~$G$ gives birth to two vertices~$x_i$
and~$\bar{x}_i$ in~$V_{2SAT}$, and every pair of adjacent
links~\mbox{$l_i \vee l_j$} in~$P$, where literal~$l_i$~($l_j$) is
either~$x_i$~($x_j$) or~$\bar{x}_i$~($\bar{x}_j$), gives birth to
two directed edges in~$E_{2SAT}$: from vertex~$\bar{l}_i$ to
vertex~$l_j$ and from vertex~$\bar{l}_j$ to vertex~$l_i$.
As shown
in~\cite{AsPlaTa79}, there exists an assignment satisfying all the
clauses if there is no edge~$i$ such that both of its
corresponding vertices in the 2SAT graph,
\mbox{$x_i,\bar{x}_i \in V_{2SAT}$}, belong to the same strongly
connected component\footnote{An SCC is a set of nodes in a
directed graph s.~t.~there exists a directed path between every
ordered pair of nodes.}~(SCC) in~$G_{2SAT}$.

If an assignment satisfying all the clauses exists we can easily
find it. We perform topological sorting\footnote{Given a directed
graph~$G(V,E)$, function~\mbox{$t:V\mapsto\mathbb{R}$} is
topological sorting if \mbox{$t(i) \leq t(j)$} for every ordered
pair of nodes \mbox{$i,j \in V$} s.~t.~there exists a directed
path from~$i$ to~$j$.}~$t$ on nodes in~$V_{2SAT}$ and assign {\em
true\/} or {\em false\/} to a variable~$x_i$ depending on
if~\mbox{$t(\bar{x}_i)<t(x_i)$} or~\mbox{$t(x_i)<t(\bar{x}_i)$}
respectively. All operations described so far can be done in
linear time.

\begin{table}[t]
  \centering
  \caption{\scriptsize Mapping between pairs of adjacent edges in~$P$,
  2SAT clauses, and edges in~$G_{2SAT}$. The invalid path pattern
  is in the last row.}
  \label{table:mapping}

  \begin{tabular}{|c|c|c|}

    \hline

    Edges in $P$ & 2SAT clause & Edges in $G_{2SAT}$ \\

    \hline

    \raisebox{2ex}[0pt]{
    \epsfig{file=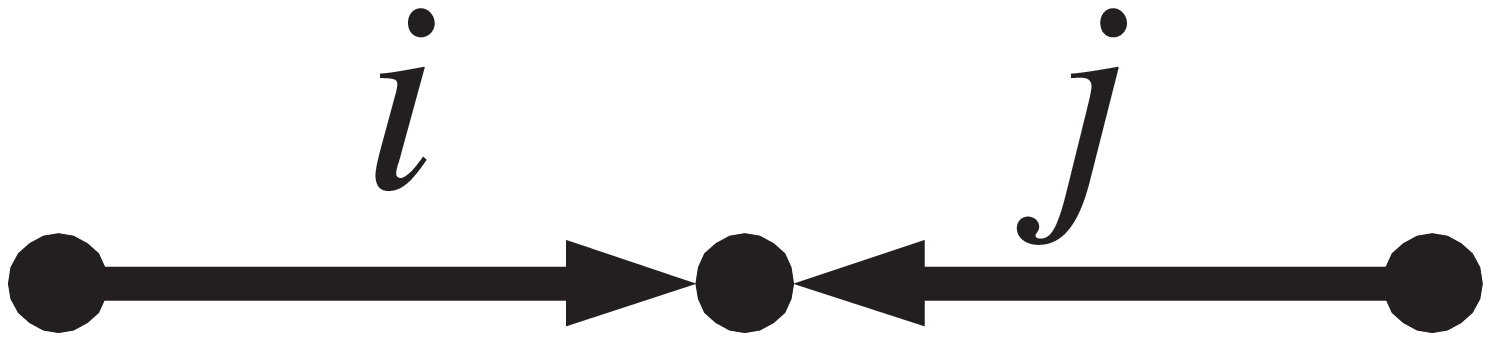,width=1in}} &
    \raisebox{3ex}[0pt]{$x_i \vee x_j$} &
    \epsfig{file=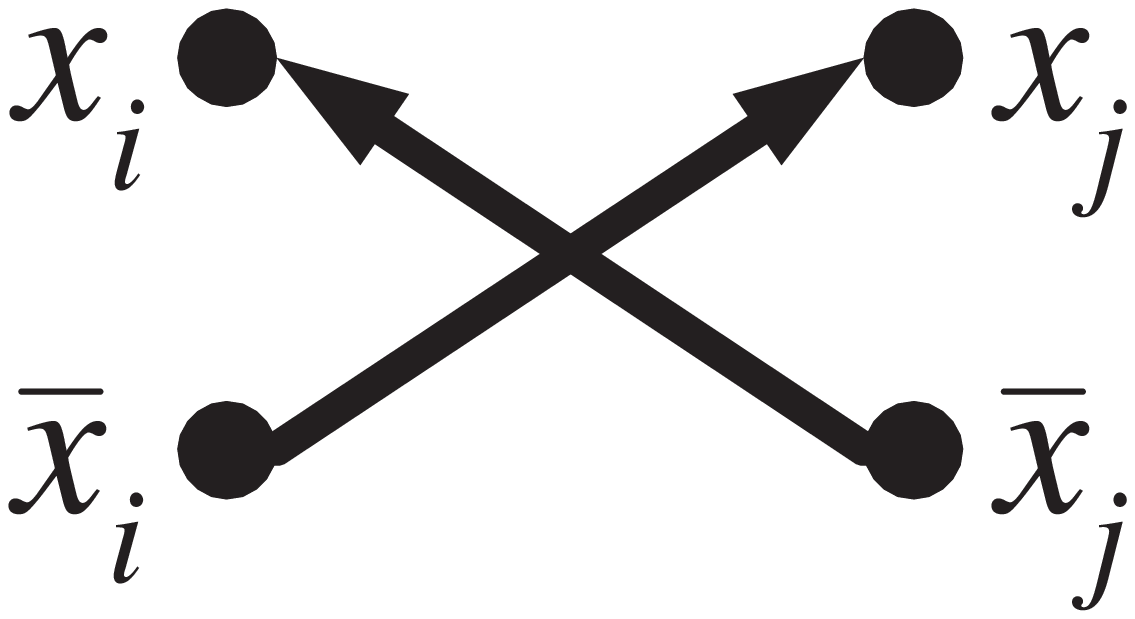,width=.75in} \\ \hline

    \raisebox{2ex}[0pt]{
    \epsfig{file=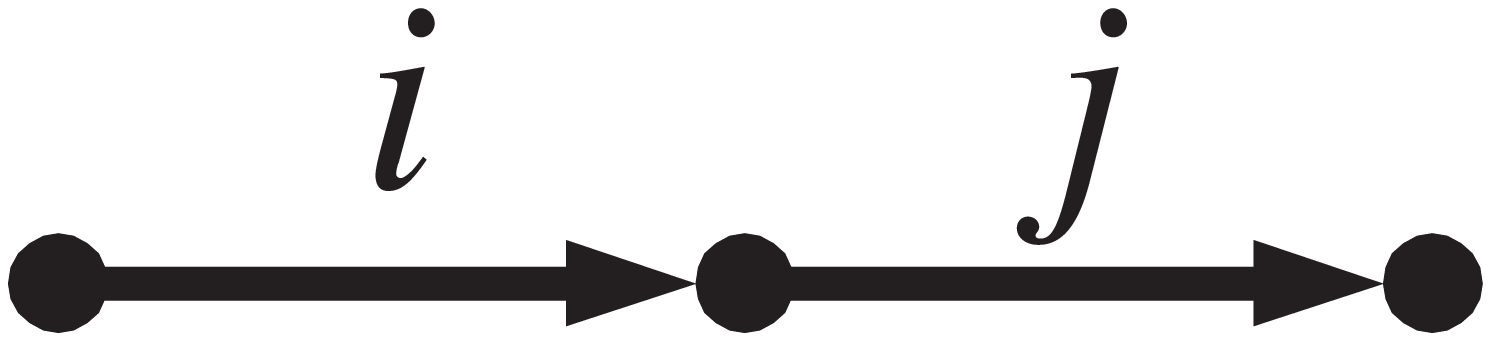,width=1in}} &
    \raisebox{3ex}[0pt]{$x_i \vee \bar{x}_j$} &
    \epsfig{file=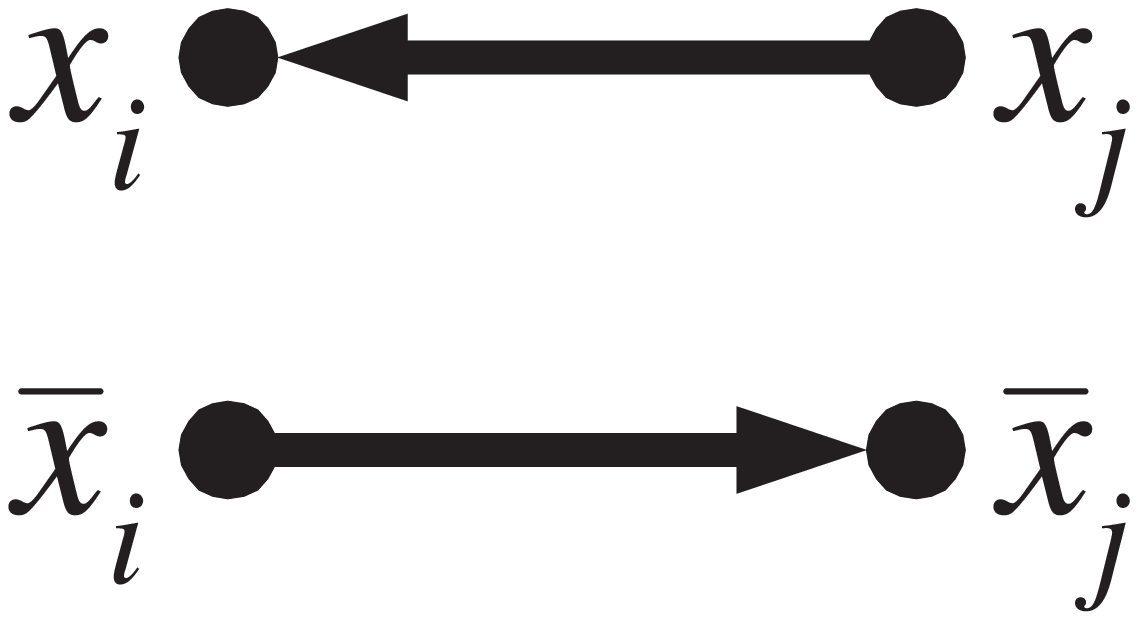,width=.75in} \\ \hline

    \raisebox{2ex}[0pt]{
    \epsfig{file=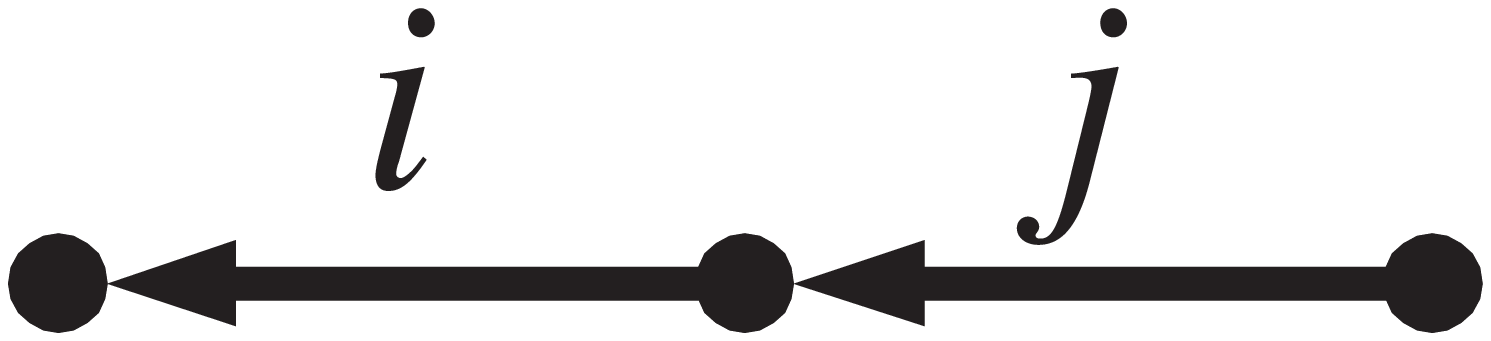,width=1in}} &
    \raisebox{3ex}[0pt]{$\bar{x}_i \vee x_j$} &
    \epsfig{file=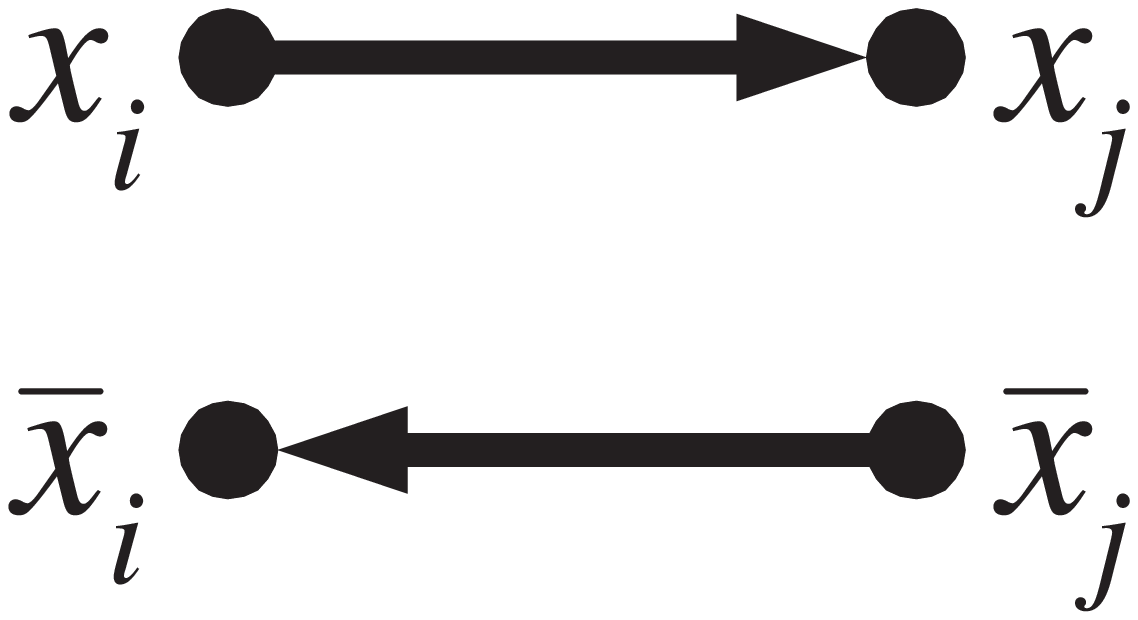,width=.75in} \\ \hline

    \raisebox{2ex}[0pt]{
    \epsfig{file=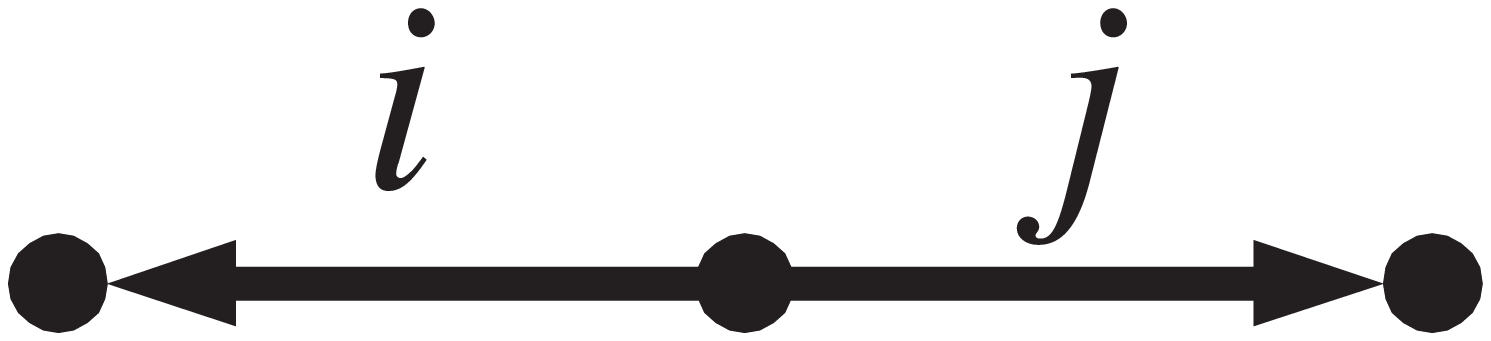,width=1in}} &
    \raisebox{3ex}[0pt]{$\bar{x}_i \vee \bar{x}_j$} &
    \epsfig{file=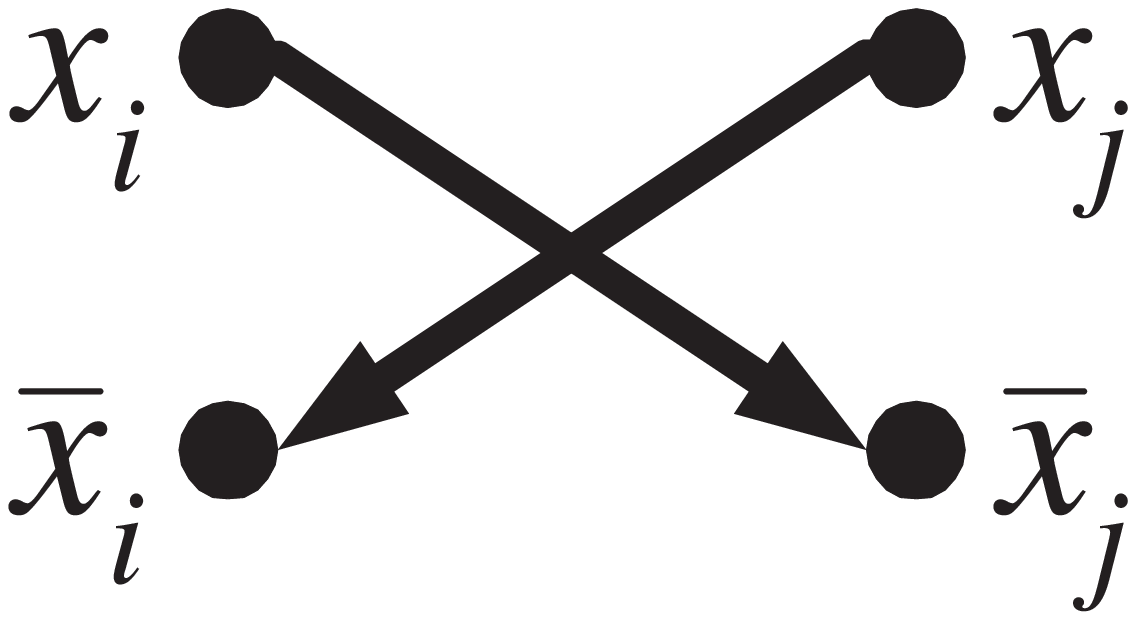,width=.75in} \\ \hline

  \end{tabular}
\end{table}

\subsection{MAX2SAT: DPP vs.~EHS}

As soon as a set of BGP paths~$P$ is ``rich enough,'' there is
no assignment satisfying all clauses and making all paths valid.
Furthermore, the ToR problem of maximizing the number of
valid paths can be reduced to the
MAX2SAT~\cite{DiBaPaPi03,ErHaSch02} problem of maximizing the
number of satisfied clauses.
Making this observation, DPP propose a heuristic to find the
maximal subset of paths~\mbox{$P_S \subset P$} such that all
paths in~$P_S$ are valid.

EHS use a different approach. They first direct the edges \mbox{$i
\in E$} that can be directed without causing conflicts. Such edges
correspond to vertices~\mbox{$x_i,\bar{x}_i \in V_{2SAT}$} that
have indegree or outdegree zero. Then EHS iteratively remove
edges directed as described above and strip~$P$, $G$,
and~$G_{2SAT}$ accordingly. This procedure significantly shortens
the average path length in~$P$, which improves the approximation
of ToR by MAX2SAT. Finally, they approximate MAX2SAT to find
a solution to the ToR problem.

\subsection{Solving MAX2SAT with SDP}

The MAX2SAT problem is NP- and
APX-complete~\cite{AuCreGaKaMaSpaPro99}, but Goemans and
Williamson (GW)~\cite{GoWi95} construct a famous approximation
algorithm that uses semidefinite programming~(SDP) and delivers
an approximation ratio of~$0.878$. The  best approximation ratio
currently known is~$0.940$, due to improvements to~GW by
Lewin, Livnat, and Zwick~(LLZ) in~\cite{LeLiZwi02}. Note that this
approximation ratio is pretty close to the MAX2SAT
inapproximability limit
of~$\frac{21}{22}\sim0.954$~\cite{hastad97}.

To cast a MAX2SAT problem with~$m_2$ clauses involving~$m_1$
literals (variables~$x_i$ and their negations~$\bar{x}_i$,
\mbox{$i=1 \ldots m_1$}) to a semidefinite program, we first get
rid of negated variables by introducing~$m_1$
variables~\mbox{$x_{m_1+i}=\bar{x}_i$}. Then we establish mapping
between boolean variables~$x_k$, \mbox{$k=1 \ldots 2m_1$},
and~\mbox{$2m_1+1$} auxiliary variables~\mbox{$y_0,y_k \in
\{-1,1\}$}, \mbox{$y_{m_1+i}=-y_i$}, using formula
\mbox{$x_k=(1+y_0y_k)/2$}. This mapping guarantees that
\mbox{$x_k=true$} $\Leftrightarrow$ \mbox{$y_k=y_0$} and
\mbox{$x_k=false$} $\Leftrightarrow$ \mbox{$y_k=-y_0$}. Given the
described construction, we call~$y_0$ the {\em truth\/} variable.
After trivial algebra, the MAX2SAT problem becomes the
maximization problem for the sum
\mbox{1/4$\sum_{k,l=1}^{2m_1}w_{kl}(3+y_0y_k+y_0y_l-y_ky_l)$},
where weights~$w_{kl}$ are either~$1$ if clause \mbox{$x_k \vee
x_l$} is present in the original MAX2SAT instance or~$0$
otherwise. Hereafter we fix the notations for indices \mbox{$i,j =
1 \ldots m_1$} and \mbox{$k,l = 1 \ldots 2m_1$}.

The final transformation to make the problem solvable by SDP is
relaxation. Relaxation involves mapping variables~$y_0,y_k$
to~$2m_1+1$ unit vectors~\mbox{$v_0,v_k \in \mathbb{R}^{m_1+1}$}
fixed at the same origin---all vector ends lie on the unit
sphere~$S_{m_1}$. The problem is to maximize the sum composed of
vector scalar products:
\begin{eqnarray}
    \max && \frac{1}{4}\sum_{k,l=1}^{2m_1} w_{kl}
        (3 + v_0 \cdot v_k + v_0 \cdot v_l - v_k \cdot v_l)
        \label{eq:obj_func}\\
    \text{s.t.} && v_0 \cdot v_0 = v_k \cdot v_k = 1, \;
        v_i \cdot v_{m_1+i} = -1, \nonumber \\
    && k = 1 \ldots 2m_1, \; i = 1 \ldots m_1. \nonumber
\end{eqnarray}

\begin{figure}[t]

\centerline{\epsfig{file=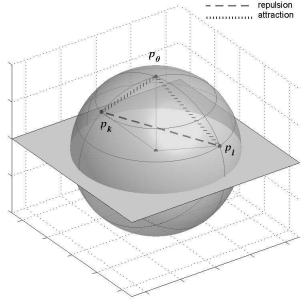,width=2.5in}}

\caption{\scriptsize The semidefinite programming relaxation to the MAX2SAT
problem. Point~$p_0$ (corresponding to vector~$v_0$ from the text)
is the {\em truth\/} point. It attracts both points~$p_k$
and~$p_l$ representing the boolean variables from the
clause~\mbox{$x_k \vee x_l$}. Points~$p_k$ and~$p_l$ repel each
other. The problem is to identify the locations of all points on
the sphere that minimize the potential energy of the system. Given
an orientation by SDP, we cut the system by a random hyperplane
and assign value {\em true\/} to the variables corresponding to
points lying on the same side of the hyperplane as the
truth~$p_0$.}

\label{fig:sphere}

\end{figure}

Interestingly, this problem, solvable by SDP, is equivalent to the
following minimum energy problem in physics. Vectors~$v_0,v_k$
point to the locations of particles~$p_0,p_k$ freely moving on the
sphere~$S_{m_1}$ except that particles~$p_i$ and~$p_{m_1+i}$ are
constrained to lie opposite on the sphere. For every MAX2SAT
clause \mbox{$x_k \vee x_l$}, we introduce three constant forces
of equal strength (see~Fig.~\ref{fig:sphere}): one repulsive force
between particles~$p_k$ and~$p_l$, and two attractive forces:
between~$p_k$ and~$p_0$, and between~$p_l$ and~$p_0$---the {\em
truth\/} particle~$p_0$ attracts all other particles~$p_k$ with
the forces proportional to the number of clauses containing~$x_k$.
The goal is to find the location of particles on the sphere
minimizing the potential energy of the system. If we built such a
mission-specific computer in the lab, it would solve this problem
in constant time. SDP solves it in polynomial time.

To extract the solution for the MAX2SAT problem from the solution
obtained by SDP for the relaxed problem, we perform rounding.
Rounding involves cutting the sphere by a randomly oriented
hyperplane containing the sphere center. We assign value {\em
true\/} ({\em false}) to variables~$x_k$ corresponding to
vectors~$v_k$ lying on the same (opposite) side of the hyperplane
as the {\em truth\/} vector~$v_0$. GW prove that the solution to
the MAX2SAT problem obtained this way delivers the approximation
ratio of~$0.878$~\cite{GoWi95}. We can also rotate the vector
output obtained by SDP before rounding and skew the distribution
of the hyperplane orientation to slightly prefer the orientation
perpendicular to~$v_0$. These two techniques explored to their
greatest depths by~LLZ improve the approximation ratio up
to~$0.9401$~\cite{LeLiZwi02}.

\subsection{Analysis of the unperturbed solution}

We now have the solution to the original ToR problem and are
ready to analyze it. While the number of invalid paths is
small~\cite{ErHaSch02}, the solution is not perfect---some
inferred AS relationships are not in fact accurate.
What causes these misclassifications?

First, some edges may be directed either way resulting in exactly
the same number of invalid paths---such edges are directed
randomly. To exemplify, consider path~\mbox{$p \in P$},
\mbox{$p=\{i_1 i_2 \ldots i_{|p|-1} j\}$},
\mbox{$i_1,i_2,\ldots,i_{|p|-1},j \in E$}, and suppose that the
last edge~$j$ appears only in one path (that is,~$p$) and that it
is from some large provider (like UUNET) to a small customer.
Suppose that other edges \mbox{$i_1,i_2,\ldots,i_{|p|-1}$} appear
in several other paths and that they are correctly inferred as
customer-to-provider. In this scenario both orientations of
edge~$j$ (i.e.~correct and incorrect: provider-to-customer and
customer-to-provider) render path~$p$ valid. Thus, edge~$j$ is
directed randomly, increasing the likelihood of an incorrect
inference. We can find many incorrect inferences of this type in
our experiments in the next section and in~\cite{rimondini02},
e.g.~well-known large providers like UUNET, AT\&T, Sprintlink,
Level3, are inferred as customers of smaller ASs like AS1~(AS
degree 67), AS2685~(2), AS8043~(1), AS13649~(7), respectively.

Second, not all edges are customer-to-provider or
provider-to-customer. In particular, trying to direct sibling
edges leads to proliferation of error. Indeed, when the only
objective is to maximize the number of valid paths, directing a
sibling edge brings the risk of misdirecting the dependent edges
sharing a clause with the sibling edge. To clarify, 
consider path~\mbox{$p \in P$},
\mbox{$p=\{ij\}$}, \mbox{$i,j \in E$}, and suppose that in
reality $i$ is a sibling edge that appears in multiple paths and
that~$j$ is a customer-to-provider edge that appears only in one
path~$p$. The algorithm can classify edge~$i$ either as
customer-to-provider or provider-to-customer depending on the
structure of the paths in which it appears. If this structure
results in directing~$i$ as provider-to-customer, then the
algorithm erroneously directs edge~$j$ also as
provider-to-customer to make path~$p$ valid. In other words, the
outcome is that we maximize the number of valid paths at the cost
of inferring edge~$j$ incorrectly.

We can conclude that the maximum number of valid paths does not
correspond to a correct answer because, as illustrated in the
above two examples, it can result in miss-inferred links.
Specifically, in the presence of multiple solutions there is
nothing in the objective function to require the algorithm
to prefer the proper orientation for edge~$j$.
Our next key question is: Can we adjust the objective function to
infer the edge direction correctly?

\subsection{Our new generalized objective function}

A rigorous way to pursue the above question is to add to the
objective function some small modifier selecting the correct edge
direction for links unresolved by the unperturbed objective
function. Ideally this modifier should be a function of ``AS
importance,'' such as the relative size of the customer tree of an
AS. Unfortunately, defined this way the modifier is a function of
the end result, edge orientation, which makes the problem
intractable (i.e.~we cannot solve it until we solve it).

The simplest correcting function that does not depend on the
edge direction and is still related to perceived ``AS
importance,'' is the AS degree ``gradient'' in the original
undirected graph~$G$---the difference between node degrees of
adjacent ASs. In the examples from the previous subsection, the
algorithm that is trying not only to minimize the number of invalid
paths but also to direct edges from adjacent nodes of lower degrees to
nodes of higher degrees will effectively have an incentive
to correctly infer the last edge~\mbox{$j \in p$}.

More formally, we modify the objective function as follows. In the
original problem formulation, weights~$w_{kl}$ for 2-link
clauses~\mbox{$x_k \vee x_l$} (pairs of adjacent links in~$P$) are
either~$0$ or~$1$. We first alter them to be either~$0$, if
pair~\mbox{$\{kl\} \notin P$},
or~\mbox{$w_{kl}(\alpha)=c_2\alpha$} otherwise. The normalization
coefficient~$c_2$ is determined from the condition \mbox{$\sum_{k
\neq l} w_{kl}(\alpha)=\alpha$} $\Rightarrow$ \mbox{$c_2 = 1/m_2$}
(recall that~$m_2$ is the number of 2-link clauses), and~$\alpha$
is an external parameter, \mbox{$0 \leq \alpha \leq 1$}, whose
meaning we explain below.

In addition, for every edge~\mbox{$i \in E$}, we introduce a
1-link clause weighted by a function of the node degree gradient.
More specifically, we initially orient every edge~\mbox{$i \in E$}
along the node degree gradient: if~$d_i^-$ and~$d_i^+$,
\mbox{$d_i^- \leq d_i^+$}, are degrees of nodes adjacent to
edge~$i$, we direct~$i$ from the $d_i^-$-degree node to the
$d_i^+$-degree node, for use as input to our
algorithm.\footnote{An initial direction along the node
degree gradient does not affect the solution since any initial
direction is possible. We select the node degree gradient direction
to simplify stripping of non-conflict edges in the next section.}

Then, we add 1-link clauses~\mbox{$x_i \vee x_i$}, \mbox{$\forall
i \in E$}, to our MAX2SAT instance, and we weight them by
\mbox{$w_{ii}(\alpha) = c_1(1-\alpha)f(d_i^-,d_i^+)$}. The
normalization coefficient~$c_1$ is determined from the condition
\mbox{$\sum_iw_{ii}(\alpha)=1-\alpha$}, and the function~$f$
should satisfy the following two conditions: 1)~it should
``roughly depend'' on the {\em relative\/} node degree
gradient~\mbox{$(d_i^+-d_i^-)/d_i^+$}; and 2)~it should provide
higher values for node pairs with the same relative degree
gradient but higher absolute degree values. The first condition is
transparent: we expect that an AS with node degree~$5$, for
example, is more likely a customer of an AS with node degree~$10$
than a $995$-degree AS is a customer of a $1000$-degree AS. The
second condition is due to the fact that we do not know the
true AS degrees: we approximate them by degrees of nodes in our
BGP-derived graph~$G$. The graphs derived from BGP data have
a tendency to underestimate the node degree of small ASs, while they
yield more accurate degrees for larger ASs~\cite{ChaGoJaSheWi04}.
Because of the larger error associated with small ASs, an AS with
node degree~$5$, for example, is less likely a customer of an AS
with node degree~$10$ than a $500$-degree AS is a customer of a
$1000$-degree AS.

We select the following function satisfying the two criteria
described above:
\begin{equation}
    f(d_i^-,d_i^+)=\frac{d_i^+-d_i^-}{d_i^++d_i^-}\log(d_i^++d_i^-).
\end{equation}

In summary, our new objective function looks exactly as
the one in~(\ref{eq:obj_func}), but with different weights on
clauses:
\begin{equation}\label{eq:weights}
    w_{kl}(\alpha) =
    \begin{cases}
        c_2\alpha                   & \text{if \; $\{kl\} \in P$,}\\
        c_1(1-\alpha)f(d_k^-,d_k^+) & \text{if \; $k=l \leq m_1$,}\\
        0                           & \text{otherwise.}
    \end{cases}
\end{equation}

Now we can explain the role of the parameter~$\alpha$.
Since~\mbox{$\sum_{k \neq l}w_{kl}(\alpha) = \alpha$}
and~\mbox{$\sum_{k=l}w_{kl}(\alpha) = 1-\alpha$},
parameter~$\alpha$ measures the relative importance of sums of all
2- and 1-link clauses. If~\mbox{$\alpha=1$}, then the problem is
equivalent to the original unperturbed ToR problem---only the
number of invalid paths matters. If~\mbox{$\alpha=0$}, then,
similar to Gao, only node degrees matter. Note that in the
terminology of multiobjective optimization, we
consider the simplest scalar method of weighted sums.

In our analogy with physics in~Fig.~\ref{fig:sphere}, we have
weakened the repulsive forces among particles other than the {\em
truth\/} particle~$p_0$, and we have strengthened the forces
between~$p_0$ and other particles. When~\mbox{$\alpha=0$}, there
are no repulsive forces, the {\em truth\/} particle~$p_0$ attracts
all other particles to itself, and all the vectors become
collinear with~$v_0$. Cut by any hyperplane, they all lie on the
same side as~$v_0$, which means that all variables~$x_i$ are
assigned value {\em true\/} and all links~$i$ remain directed
along the node degree gradient in the output of our algorithm.

\section{Results}
\begin{figure}[t]

\centerline{\epsfig{file=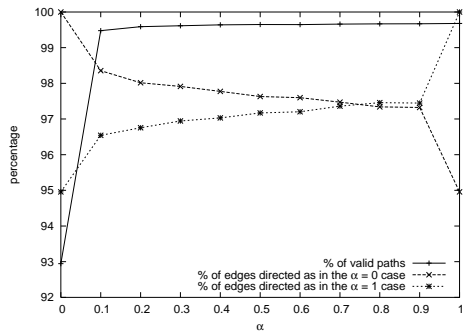,width=2.5in}}

\caption{\scriptsize Percentage of valid paths, of edges directed as
    in the $\alpha = 0$ case and of edges directed as in
    the $\alpha = 1$ case for different values of~$\alpha$.}

\label{fig:results}

\end{figure}

In our experiments, the BGP path set~$P$ is a union of BGP tables
from RouteViews~\cite{routeviews} and 18 BGP route servers
from~\cite{traceroute.org} collected on May 13, 2004. Paths of
length~1 are removed since they are always valid. The total number
of paths is~1,025,775 containing 17,557 ASs, 37,021 links, and 382,917
unique pairs of adjacent links.

We first pre-process the data by discovering sibling links. For
this purpose, we use a union of WHOIS databases from ARIN, RIPE,
APNIC, and LACNIC collected on June 10, 2004. We say that two ASs
belong to the same organization if, in the WHOIS database, they
have exactly the same organization names, or names different only
in the last digits, e.g.~\mbox{``ATT-37''} and
\mbox{``ATT-38,''}
or very similar names, e.g.~``UUNET South
Africa'' and ``UUNET Germany.'' We infer links in~$P$ between
adjacent ASs belonging to the same organization as sibling. We
find 211 sibling links in our dataset, which we ignore in
subsequent steps. More precisely, we do not assign boolean
variables to them.

We then direct the remaining links in the original graph~$G$ along
the node degree gradient, assign boolean variables to them, and
construct the dual~$G_{2SAT}$ graph. After directing edge~$i$
along the node degree gradient, we check whether this direction
satisfies all clauses containing~$l_i$ ($x_i$ or~$\bar{x}_i$). If
so, we then remove the edge and strip~$P$, $G$, and~$G_{2SAT}$
accordingly. In this case we say that edge~$i$ causes no conflicts
because the value of the corresponding literal~$l_i$ satisfies all
the clauses in which~$l_i$ appears, independent of the values of
all other literals sharing the clauses with~$l_i$. A non-conflict
edge has two corresponding vertices in the $G_{2SAT}$ graph, $x_i$
and~$\bar{x}_i$. It follows from the construction of the
$G_{2SAT}$ graph that~$x_i$ has an outdegree of zero
and~$\bar{x}_i$ has an indegree of zero. We repeat the described
procedure until we cannot remove any more edges. The stripped
graph~$G$ has 1,590 vertices (9\% of the original~$\big|V\big|$)
and 4,249 edges (11\% of the original~$\big|E\big|$). The
stripped~$G_{2SAT}$ graph has 8,498 vertices and 46,920 edges. In
summary, we have 4,249~($m_1$) 1-link clauses and 23,460~($m_2$)
2-link clauses. We feed this data into a publicly available SDP
solver DSDP~v4.7~\cite{dsdp}, reusing parts of the code
from~\cite{ErHaSch02} and utilizing the LEDA~v4.5 software
library~\cite{leda}. We incorporate the pre-rounding rotation and
skewed distribution of hyperplane orientation by
LLZ~\cite{LeLiZwi02}.

Fig.~\ref{fig:results} shows results of edge orientations we
derive for different values of~$\alpha$ in~(\ref{eq:weights}).
Specifically, the figure shows the percentage of valid paths,
edges directed as in the \mbox{$\alpha=0$} case, and edges
directed as in the \mbox{$\alpha=1$} case. In the particular
extreme case of~\mbox{$\alpha=1$}, the problem reduces to the
original ToR problem considered by DPP and EHS, and its
solution yields the highest percentage of valid paths,~$99.67\%$. By
decreasing~$\alpha$, we increase preference to directing edges
along the node degree gradient, and at the other extreme
of~\mbox{$\alpha=0$}, all edges become directed along
the node gradient, but the number of valid
paths is~92.95\%.

\begin{table*}[ct]
    \centering
    \caption{\scriptsize
    Hierarchical ranking of ASs. The position {\em depth\/} (the number of AS at the
    levels above) and {\em width\/} (the number of ASs at the same level)
    of the top five ASs in the \mbox{$\alpha = 0$} and
    \mbox{$\alpha = 1$} cases are shown for different values of~$\alpha$.
    The customer leaf ASs are marked with asterisks.
    }
    \label{table:ranking}
\begin{tabular}{|c|c|c|cc|cc|cc|cc|cc|}
\hline
\multicolumn{3}{|c|}{ }
&\multicolumn{2}{c|}{ $\alpha = 0.0 $} &\multicolumn{2}{c|}{ $\alpha = 0.2 $} &\multicolumn{2}{c|}{ $\alpha = 0.5 $} &\multicolumn{2}{c|}{ $\alpha = 0.8 $} &\multicolumn{2}{c|}{ $\alpha = 1.0 $} \\ \hline
AS \# &  name & degree & dep. & wid. & dep. & wid. & dep. & wid. & dep. & wid. & dep. & wid. \\ \hline
\hline
 701 & UUNET & 2373  & 0&1  & 0&173  & 1&232  & 1&252  & 17&476 \\ \hline 
 1239 & Sprint & 1787  & 1&1  & 0&173  & 1&232  & 1&252  & 17&476 \\ \hline 
 7018 & AT\&T & 1723  & 2&1  & 0&173  & 1&232  & 1&252  & 17&476 \\ \hline 
 3356 & Level 3 & 1085  & 3&1  & 0&173  & 1&232  & 1&252  & 17&476 \\ \hline 
 209 & Qwest & 1072  & 4&1  & 0&173  & 1&232  & 1&252  & 17&476 \\ \hline 
\hline
 3643 & Sprint Austr. & 17  & 194&1  & 222&1  & 250&1  & 268&1  & 0&4 \\ \hline 
 6721 & Nextra Czech Net & 3  & 1742&941  & 833&88  & 868&90  & 884&89  & 0&4 \\ \hline 
 11551 & Pressroom Ser. & 2  & 1742&941  & 1419&398  & 1445&390  & 1457&386  & 0&4 \\ \hline 
 1243 & Army Systems & 2  & 2683&14725{\small *}  & 2753&14655{\small *}  & 1445&390  & 1457&386  & 0&4 \\ \hline 
 6712 & France Transpac & 2  & 2683&14725{\small *}  & 2753&14655{\small *}  & 292&3  & 1&252  & 4&13 \\ \hline 
\end{tabular}
\end{table*}

Note that changing~$\alpha$ from~0 to~0.1 redirects 1.64\% of
edges, which leads to a significant 6.53\% increase in the number of
valid paths. We also observe that the tweak of~$\alpha$ from~1
to~0.9 redirects~2.56\% of edges without causing any significant
decrease (only 0.008\%) in the number of valid paths. We find that
most of these edges are directed randomly in the~\mbox{$\alpha=1$}
case because oriented either way they yield the same number of
valid paths. In other words, the AS relationships represented by
these edges cannot be inferred by minimizing the number of invalid
paths.

We also rank ASs by means of our inference results with
different~$\alpha$ values. To this end we split all ASs into
hierarchical levels as follows. We first order all ASs by their
{\em reachability}---that is, the number of ASs that a given AS
can reach ``for free'' traversing only provider-to-customer edges.
We then group ASs with the same reachability into levels. ASs at
the highest level can reach all other ASs ``for free.'' ASs at the
lowest level have the smallest reachability (fewest ``free''
destinations). Then we define the position {\em depth\/} of AS~X
as the number of ASs at the levels above the level of AS~X. The
position {\em width\/} of AS~X is the number of ASs at the same
level as AS~X.

Table~\ref{table:ranking} shows the results of our AS ranking. For
different values of~$\alpha$, we track the positions of the top
five ASs in the~\mbox{$\alpha=0$} and~\mbox{$\alpha=1$} cases. In
the former case, well-known large ISPs are at the
top, but the number of invalid paths is relatively large,
cf.~Fig.~\ref{fig:results}. In the latter case delivering the
solution to the unperturbed ToR problem, ASs with small degrees
occupy the top positions in the hierarchy. These ASs appear in
much lower positions when~\mbox{$\alpha\neq1$}. Counter to
reality, the large ISPs are not even near the top of the
hierarchy. We observe that the depth\footnote{Note that the large
ISPs are at the same depth as soon as~\mbox{$\alpha\neq0$}, which
is expected since they form ``almost a clique''~\cite{SuAgReKa02}
and are likely to belong to the same SCC. All nodes in the same
SCC have the same reachability. The converse is not necessarily
true.} of these large ASs increases as~$\alpha$ approaches~1,
indicating an increasingly stronger deviation from reality. The
deviation is maximized when~\mbox{$\alpha=1$}. This observation
pronounces the limitation of the ToR problem formulation based
solely on maximization of the number of valid paths.

\section{Conclusion and future work}

Using a standard multiobjective optimization method, we have
constructed a natural generalization of the known AS relationship
inference heuristics. We have extended the combinatorial
optimization approach based on minimization of invalid paths, by
incorporating AS-degree-based information into the problem
formulation. Utilizing this technique, we have obtained first
results that are more realistic than the inferences produced by
the recent
state-of-the-art heuristics~\cite{DiBaPaPi03,ErHaSch02}. We
conclude that our approach opens a promising path toward
increasingly veracious inferences of business relationships
between ASs.

The list of open issues that we plan to address in our future work
includes: 1)~modifications to the algorithm to infer {\em
peering}; 2)~careful analysis of the {\em trade-off surface}~\cite{CoSi03}
of the problem, required for
selecting the value of the external parameters (e.g.~$\alpha$)
corresponding to the right answer; 3)~detailed examination of
the {\em structure\/} of the AS graph directed by inferred AS
relationships; 4){\em~validation\/} considered as a set of
constraints narrowing the range of feasible values of external
parameters; and 5)~investigation of other {\em AS-ranking mechanisms}
responsible for the structure of the inferred AS hierarchy.

\section{Acknowledgements}

We thank Thomas Erlebach, Alexander Hall and Thomas Schank
for sharing their code with us.

Support for this work was provided by the Cisco University Research
Program, by DARPA N66002-00-1-893, and by NSF ANI-0221172, ANI-9977544,
ANI-0136969 and CNS-0434996, with support from DHS/NCS.

\bibliographystyle{splncs}
\bibliography{bib}

\end{document}